\documentclass{PoS}

\pdfoutput=1

\usepackage{amsmath}
\usepackage{here}
\usepackage{graphicx}
\usepackage{epsfig}
\usepackage{subcaption}
\usepackage{cite}
\usepackage{mathtools}
\usepackage{bm}
\usepackage{enumitem}

\title{The Loop-Tree Duality: Progress Report}

\ShortTitle{The Loop-Tree Duality: Progress Report}

\author{\speaker{Grigorios Chachamis}\\
        Instituto de F{\' \i}sica Te{\' o}rica UAM--CSIC, Nicol{\'a}s Cabrera 15\\
         E-28049 Madrid, Spain.\\
        E-mail: \email{chachamis@gmail.com}}
        
\author{Germ\'an Rodrigo\\
Instituto de F\'{\i}sica Corpuscular, Universitat de Val\`{e}ncia 
-- Consejo Superior de Investigaciones Cient\'{\i}ficas, 
Parc Cient\'{\i}fic, E-46980 Paterna, Valencia, Spain.\\
E-mail: \email{german.rodrigo@csic.es}}

\abstract{We review the recent developments of the Loop-Tree Duality method, focussing our discussion on the first numerical implementation and its use in the direct numerical computation of multi-leg Feynman integrals. Non-trivial examples are presented.}

\FullConference{ 25th International Workshop on Deep Inelastic Scattering and Related Topics \\
                 3-7 April 2017 \\
                 University of Birmingham, Birmingham, UK}

\begin{document}

\section{Introduction}

The Loop-Tree Duality (LTD) method~\cite{Catani:2008xa,Bierenbaum:2010cy,Bierenbaum:2012th,Bierenbaum:2013nja,Buchta:2014dfa,Buchta:2014fva,Buchta:2015vha,Sborlini:2015uia,Buchta:2015wna,Buchta:2016wfg,
Hernandez-Pinto:2015ysa,Sborlini:2016gbr,Sborlini:2016hat,Chachamis:2016olm}
turns $N$-leg loop quantities (integrals and amplitudes) into 
a sum of connected tree-level-like diagrams 
with a remaining integration measure that 
 is similar to the  $(N+1)$--body phase-space~\cite{Catani:2008xa}. 
Therefore,  loop and tree-level  corrections of the same order, may in principle be 
 treated under a common integral sign
with the use of a proper numerical integrator (usually a Monte Carlo routine)~\cite{Hernandez-Pinto:2015ysa,Sborlini:2016gbr}.
The LTD method fits into a broader effort to produce fully automated next-to-leading order (NLO) computations.
Many steps toward that direction have been taken in the last years~\cite{Soper:1998ye,Soper:1999xk,Soper:2001hu,Kramer:2002cd,Ferroglia:2002mz,Nagy:2003qn,Nagy:2006xy,Moretti:2008jj,Gong:2008ww,Kilian:2009wy,Becker:2010ng,Binoth:2010nha,Becker:2012aqa,DeRoeck:2009id,AlcarazMaestre:2012vp,Becker:2012nk,Seth:2016hmv,Gnendiger:2017pys,Bevilacqua:2011xh,Cascioli:2011va,Cullen:2014yla,Frixione:2008ym,Gleisberg:2008ta,Alwall:2014hca}. 
Substantial progress has also been made at higher orders~\cite{Passarino:2001wv,Anastasiou:2007qb,Becker:2012bi}.

Here we focus on the use of the LDT framework in
computing one-loop  Feynman diagrams.  
The numerical implementation of the LTD had been initially tested  on
integrals with up to six external legs\cite{Chachamis:2016olm}. Here we report on the performance
of the method for diagrams with up to eight external legs and we present non-trivial examples
of a scalar and tensor octagon with different internal mass configurations. The motivation for the work presented here 
originated from our intention to use the method for the computation of the $N$-photon amplitude ($2 \gamma \rightarrow (N-2) \gamma$)\cite{Mahlon:1993fe,Nagy:2006xy,Gong:2008ww,Ossola:2007bb,Bernicot:2007hs,Binoth:2007ca}.

\section{ Numerical Implementation of the LDT}

In dimensional regularisation, a one-loop scalar diagram can be represented by 
\begin{equation}
L^{(1)}(p_1, p_2,\dots , p_N) =  -i\int \frac{d^d\ell}{(2 \pi)^d} \prod\limits_{i=1}^NG_F(q_i)~,
\label{eq-a}
 \end{equation}
where $\ell = (\ell_0, {\boldsymbol{\ell}})$ is the loop momentum, $G_F(q_i) = 1/(q_i^2-m_i^2+i0)$ are Feynman propagators and
 $q_i$ are the momenta of the internal lines  which depend on  $\ell$.
 By applying the LTD, we essentially integrate over the energy component $\ell_0$
 using the residue theorem.
  The loop diagram turns then into a sum of integrals over the three-momentum $\boldsymbol{\ell}$
 each of which is called a ``dual contribution''. The dual contributions 
  emerge from the original integral after cutting one of the internal lines:
  \begin{figure}[H]
  \centering
  \begin{subfigure}[h]{0.3\textwidth}
  \begin{equation}
L^{(1)}(p_1,p_2,\dots ,p_N)= \nonumber
\end{equation}
\end{subfigure}
\begin{subfigure}[h]{0.69\textwidth}
  \includegraphics[scale=1]{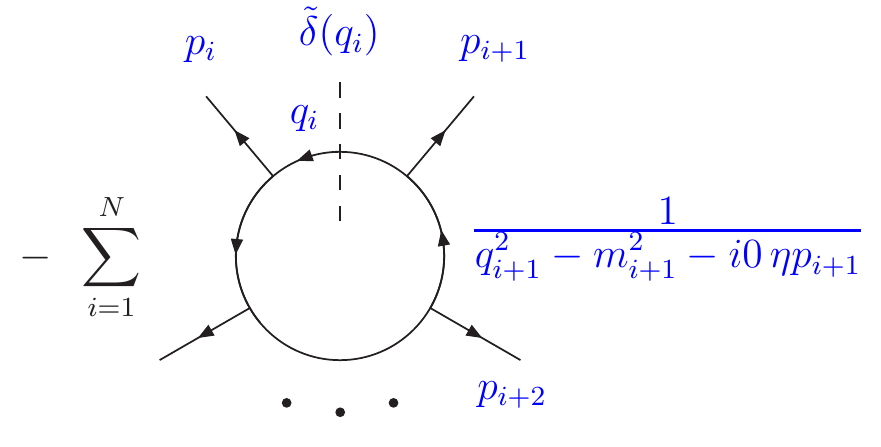}
    \end{subfigure}
  \end{figure}
 \hspace{-.85cm}
  where $\tilde{\delta}(q_i)=2\pi i\delta_+(q_i^2-m_i^2)$ with the ``+''  subscript stating that we are taking
   the positive-energy solution. 
To integrate the dual contributions over $\boldsymbol{\ell}$
requires most of the times a  contour deformation due to the presence of the
so-called {\it ellipsoid} and {\it hyperboloid}  singularities~\cite{Buchta:2015wna}
that in general are present at the integrand level.

The LTD method has been implemented  in a
C\texttt{++} code~\cite{Buchta:2015wna} and for the numerical integration
 the {\tt Cuba} library~\cite{Hahn:2004fe} was used.  
One needs only to provide the external four-momenta and the internal propagator masses. There is freedom
from the side of the user to change various parameters, e.g. the parameters of the contour deformation,
choose an integration routine between {\tt Cuhre}~\cite{Cuhre1, Cuhre2} and {\tt VEGAS}~\cite{Lepage:1980dq}
and  specify the desired number of evaluations or the required accuracy.
At run time, the code initially reads in and assigns masses and external momenta. Then it proceeds with 
an analysis of the ellipsoid and hyperboloid singularity structure to set up the details of the contour deformation
and finally performs the numerical integration using either {\tt Cuhre} or {\tt VEGAS}.
It has been tested for a large number of scalar and 
tensor diagrams with different number of external legs using as third-party reference values results from {\tt LoopTools 2.10}~\cite{Hahn:1998yk}
and  {\tt SecDec 3.0}~\cite{Borowka:2015mxa}. 
The running time for a precision of 4-digits,
on a typical Desktop machine  (Intel i7 @ 3.4 GHz processor, 4-cores 8-threads),
varied from below a second to around 30 seconds.

In Table~\ref{tab:hexagonlp}, we present results for a scalar and tensor octagon. The former has all internal masses different
whereas the latter has all internal masses equal.
\begin{table}[htb]
\begin{center}
\begin{tabular}{|c|c|c|} \hline
Diagram   & Real Part  &  Imaginary Part \\
\hline
       Scalar octagon       & $~~~~~6.8263(4)\times 10^{-10}$ & $~~~+ i~9.17379(37)\times 10^{-10}$ \\
\hline
      Tensor octagon       & $~~~~~-3.77449(34)\times 10^{-10}$ & $+ i~2.82760(3)\times 10^{-9}$ \\
\hline
\end{tabular}
\caption{Tensor octagon with all internal masses equal.
\label{tab:hexagonlp}}
\end{center}
\end{table}
\vspace{-.3cm}
The external momenta configuration used for both the scalar and tensor octagons is shown in Eq.~(2.2)\\
\begin{equation}
\begin{aligned}
&\hspace{-1.5cm}p_1 = (-2.500000,  \hspace{1.cm}0,   \hspace{2.2cm}        0,   \hspace{1.9cm}       -2.500000)     \nonumber   \\[-2pt] 
&\hspace{-1.5cm}p_2 = (-2.500000,  \hspace{1.cm}0,   \hspace{2.2cm}        0,   \hspace{2.2cm}         2.500000)            \\[-2pt] 
&\hspace{-1.5cm}p_3 = (-0.427656,  \hspace{1.cm}0.041109, \hspace{.65cm}  -0.180818,  \hspace{.95cm}  0.385362) \\[-2pt] 
&\hspace{-1.5cm}p_4 = (-0.907144,  \hspace{1.cm} 0.289299,   \hspace{.95cm}0.859318,  \hspace{.95cm} 2.805929) \\[-2pt] 
&\hspace{-1.5cm}p_5 = (-0.414246,  \hspace{1.cm} 0.329547, \hspace{.95cm}  0.249476, \hspace{.65cm} -0.027570)  \\[-2pt] 
&\hspace{-1.5cm}p_6 = (-1.907351, \hspace{.7cm} -0.950926, \hspace{.65cm} -1.460214,   \hspace{.95cm}0.775566)  \\[-2pt] 
&\hspace{-1.5cm}p_7 = (-0.271157,  \hspace{1.cm} 0.155665,  \hspace{.95cm} 0.039639,  \hspace{.65cm}-0.218456) \\[-2pt] 
&\hspace{-1.5cm}p_8 = - p_1 - p_2 - p_3 - p_4 - p_5 - p_6 - p_7 
\end{aligned}\,\,,
\tag{2.2} \label{eq:2.2}
\end{equation}
whereas the numerators and the masses for the two cases are given below:
\begin{enumerate}[leftmargin=3cm]
 \setlength\itemsep{.5em}
\item[\vspace{5cm}{\bf Scalar octagon}]
\text{numerator}: 1\\
masses:\\
$m_1 = 4.506760$\,,
$m_2 = 2.814908$\,,
$m_3 = 1.427626$\,,
$m_4 = 7.621541$\\
$m_5 = 5.269166$\,,
$m_6 = 3.521039$\,,
$m_7 = 5.888145$\,,
$m_8 = 4.422515$
\item[{\bf Tensor octagon}]
\text{numerator}: $\ell.p_2 \times \ell.p_4$\\
\text{masses}: $m_1 = m_2 = m_3 = m_4 = m_5 = m_6 = m_7 = m_8 = 4.506760$.
\end{enumerate}

\section{Conclusions}

The LTD method exhibits many interesting theoretical properties
 when  processes with many external 
legs and different mass scales are under consideration. Our numerical implementation
of the LTD demonstrates many of the method's appealing characteristics.
The code has an excellent performance for integrals 
with many external legs since 
it shows only a moderate rise in the running time as the number of legs increases.

Our next step will be to apply our LTD numerical implementation
on the computation of $N$-photon amplitudes. It would also be interesting although
more technically involved, to apply the LTD in processes with
$N$-gluon one-loop amplitudes demanding two of gluons to be off-shell.

\begin{flushleft}
\vspace{-.2cm}
{\bf \large Acknowledgements}
\end{flushleft}
\vspace{-.2cm}
This work has been supported by the Spanish Research Agency (Agencia Estatal de Investigación) through the grant IFT Centro de Excelencia Severo Ochoa SEV-2016-0597, by the Spanish Government and ERDF funds from the European  Commission (Grants No. FPA2014-53631-C2-1-P and SEV-2014-0398) and the Consejo Superior de Investigaciones Cientificas (Grant No. PIE-201750E021). GC acknowledges support from the MICINN, Spain, under contract FPA2016-78022-P.


\begin{thebibliography}{99}


\bibitem{Catani:2008xa}
  S.~Catani, T.~Gleisberg, F.~Krauss, G.~Rodrigo and J.~C.~Winter,
  ``From loops to trees by-passing Feynman's theorem,''
  JHEP {\bf 0809} (2008) 065
  [arXiv:0804.3170 [hep-ph]].

\bibitem{Bierenbaum:2010cy}
  I.~Bierenbaum, S.~Catani, P.~Draggiotis and G.~Rodrigo,
  ``A Tree-Loop Duality Relation at Two Loops and Beyond,''
  JHEP {\bf 1010} (2010) 073
  [arXiv:1007.0194 [hep-ph]].



\bibitem{Bierenbaum:2012th}
  I.~Bierenbaum, S.~Buchta, P.~Draggiotis, I.~Malamos and G.~Rodrigo,
  ``Tree-Loop Duality Relation beyond simple poles,''
  JHEP {\bf 1303} (2013) 025
  [arXiv:1211.5048 [hep-ph]].

\bibitem{Bierenbaum:2013nja}
  I.~Bierenbaum, P.~Draggiotis, S.~Buchta, G.~Chachamis, I.~Malamos and G.~Rodrigo,
  ``News on the loop--tree Duality,''
  Acta Phys.\ Polon.\ B {\bf 44} (2013) 2207.
  
\bibitem{Buchta:2014dfa}
  S.~Buchta, G.~Chachamis, P.~Draggiotis, I.~Malamos and G.~Rodrigo,
  ``On the singular behaviour of scattering amplitudes in quantum field theory,''
  JHEP {\bf 1411} (2014) 014
  [arXiv:1405.7850 [hep-ph]].

\bibitem{Buchta:2014fva}
  S.~Buchta, G.~Chachamis, I.~Malamos, I.~Bierenbaum, P.~Draggiotis and G.~Rodrigo,
  ``The loop-tree duality at work,''
  PoS LL {\bf 2014} (2014) 066
  [arXiv:1407.5865 [hep-ph]].

\bibitem{Buchta:2015vha}
  S.~Buchta, G.~Chachamis, P.~Draggiotis, I.~Malamos and G.~Rodrigo,
  ``Towards a Numerical Implementation of the Loop-Tree Duality Method,''
  Nucl.\ Part.\ Phys.\ Proc.\  {\bf 258-259} (2015) 33
  [arXiv:1509.07386 [hep-ph]].

\bibitem{Sborlini:2015uia} 
  G.~F.~R.~Sborlini, R.~Hernández-Pinto and G.~Rodrigo,
  ``From dimensional regularization to NLO computations in four dimensions,''
  PoS EPS {\bf -HEP2015}, 479 (2015)
  [arXiv:1510.01079 [hep-ph]].

  
\bibitem{Buchta:2015wna} 
  S.~Buchta, G.~Chachamis, P.~Draggiotis and G.~Rodrigo,
  Eur.\ Phys.\ J.\ C {\bf 77}, no. 5, 274 (2017)
  doi:10.1140/epjc/s10052-017-4833-6
  [arXiv:1510.00187 [hep-ph]].
  
\bibitem{Buchta:2016wfg} 
  S.~Buchta, G.~Chachamis, I.~Malamos, I.~Bierenbaum, P.~Draggiotis and G.~Rodrigo,
  Nucl.\ Part.\ Phys.\ Proc.\  {\bf 273-275}, 2009 (2016).
  doi:10.1016/j.nuclphysbps.2015.09.325
  
\bibitem{Hernandez-Pinto:2015ysa} 
  R.~J.~Hernandez-Pinto, G.~F.~R.~Sborlini and G.~Rodrigo,
  ``Towards gauge theories in four dimensions,''
  JHEP {\bf 1602}, 044 (2016)
  [arXiv:1506.04617 [hep-ph]].
  
\bibitem{Sborlini:2016gbr} 
  G.~F.~R.~Sborlini, F.~Driencourt-Mangin, R.~Hernandez-Pinto and G.~Rodrigo,
  ``Four-dimensional unsubtraction from the loop-tree duality,''
  arXiv:1604.06699 [hep-ph].

\bibitem{Sborlini:2016hat} 
  G.~F.~R.~Sborlini, F.~Driencourt-Mangin and G.~Rodrigo,
  JHEP {\bf 1610}, 162 (2016)
  doi:10.1007/JHEP10(2016)162
  [arXiv:1608.01584 [hep-ph]].
  
\bibitem{Chachamis:2016olm} 
  G.~Chachamis, S.~Buchta, P.~Draggiotis and G.~Rodrigo,
  PoS DIS {\bf 2016}, 067 (2016)
  [arXiv:1607.00875 [hep-ph]].
    
\bibitem{Soper:1998ye}
  D.~E.~Soper,
  ``QCD calculations by numerical integration,''
  Phys.\ Rev.\ Lett.\  {\bf 81} (1998) 2638
  [hep-ph/9804454].

\bibitem{Soper:1999xk}
  D.~E.~Soper,
  ``Techniques for QCD calculations by numerical integration,''
  Phys.\ Rev.\ D {\bf 62} (2000) 014009
  [hep-ph/9910292].

\bibitem{Soper:2001hu}
  D.~E.~Soper,
  ``Choosing integration points for QCD calculations by numerical integration,''
  Phys.\ Rev.\ D {\bf 64} (2001) 034018
  [hep-ph/0103262].

\bibitem{Kramer:2002cd}
  M.~Kramer and D.~E.~Soper,
  ``Next-to-leading order numerical calculations in Coulomb gauge,''
  Phys.\ Rev.\ D {\bf 66} (2002) 054017
  [hep-ph/0204113].

\bibitem{Ferroglia:2002mz}
  A.~Ferroglia, M.~Passera, G.~Passarino and S.~Uccirati,
  ``All purpose numerical evaluation of one loop multileg Feynman diagrams,''
  Nucl.\ Phys.\ B {\bf 650} (2003) 162
  [hep-ph/0209219].

\bibitem{Nagy:2003qn}
  Z.~Nagy and D.~E.~Soper,
  ``General subtraction method for numerical calculation of one loop QCD matrix elements,''
  JHEP {\bf 0309} (2003) 055
  [hep-ph/0308127].

\bibitem{Nagy:2006xy}
  Z.~Nagy and D.~E.~Soper,
  ``Numerical integration of one-loop Feynman diagrams for N-photon amplitudes,''
  Phys.\ Rev.\ D {\bf 74} (2006) 093006
  [hep-ph/0610028].

\bibitem{Moretti:2008jj}
  M.~Moretti, F.~Piccinini and A.~D.~Polosa,
  ``A Fully Numerical Approach to One-Loop Amplitudes,''
  arXiv:0802.4171 [hep-ph].

\bibitem{Gong:2008ww}
  W.~Gong, Z.~Nagy and D.~E.~Soper,
  ``Direct numerical integration of one-loop Feynman diagrams for N-photon amplitudes,''
  Phys.\ Rev.\ D {\bf 79} (2009) 033005
  [arXiv:0812.3686 [hep-ph]].

\bibitem{Kilian:2009wy}
  W.~Kilian and T.~Kleinschmidt,
  ``Numerical Evaluation of Feynman Loop Integrals by Reduction to Tree Graphs,''
  arXiv:0912.3495 [hep-ph].

\bibitem{Becker:2010ng}
  S.~Becker, C.~Reuschle and S.~Weinzierl,
  ``Numerical NLO QCD calculations,''
  JHEP {\bf 1012} (2010) 013
  [arXiv:1010.4187 [hep-ph]].

\bibitem{Binoth:2010nha} 
  T.~Binoth {\it et al.} [SM and NLO Multileg Working Group Collaboration],
  arXiv:1003.1241 [hep-ph].

\bibitem{Becker:2012aqa}
  S.~Becker, C.~Reuschle and S.~Weinzierl,
  ``Efficiency Improvements for the Numerical Computation of NLO Corrections,''
  JHEP {\bf 1207} (2012) 090
  [arXiv:1205.2096 [hep-ph]].

\bibitem{DeRoeck:2009id} 
  A.~De Roeck {\it et al.},
  Eur.\ Phys.\ J.\ C {\bf 66}, 525 (2010)
  doi:10.1140/epjc/s10052-010-1244-3
  [arXiv:0909.3240 [hep-ph]].

\bibitem{AlcarazMaestre:2012vp} 
  J.~Alcaraz Maestre {\it et al.} [SM and NLO MULTILEG Working Group and SM MC Working Group],
  arXiv:1203.6803 [hep-ph].

\bibitem{Becker:2012nk}
  S.~Becker and S.~Weinzierl,
  ``Direct contour deformation with arbitrary masses in the loop,''
  Phys.\ Rev.\ D {\bf 86} (2012) 074009
  [arXiv:1208.4088 [hep-ph]].

\bibitem{Seth:2016hmv} 
  S.~Seth and S.~Weinzierl,
  Phys.\ Rev.\ D {\bf 93}, no. 11, 114031 (2016)
  doi:10.1103/PhysRevD.93.114031
  [arXiv:1605.06646 [hep-ph]].

\bibitem{Gnendiger:2017pys} 
  C.~Gnendiger {\it et al.},
  Eur.\ Phys.\ J.\ C {\bf 77}, no. 7, 471 (2017)
  doi:10.1140/epjc/s10052-017-5023-2
  [arXiv:1705.01827 [hep-ph]].
  

\bibitem{Bevilacqua:2011xh} 
  G.~Bevilacqua, M.~Czakon, M.~V.~Garzelli, A.~van Hameren, A.~Kardos, C.~G.~Papadopoulos, R.~Pittau and M.~Worek,
  ``Helac-nlo,''
  Comput.\ Phys.\ Commun.\  {\bf 184}, 986 (2013)
  [arXiv:1110.1499 [hep-ph]].
  
\bibitem{Cascioli:2011va}
  F.~Cascioli, P.~Maierhofer and S.~Pozzorini,
  ``Scattering Amplitudes with Open Loops,''
  Phys.\ Rev.\ Lett.\  {\bf 108} (2012) 111601
  [arXiv:1111.5206 [hep-ph]].

\bibitem{Cullen:2014yla} 
  G.~Cullen {\it et al.},
  ``G$\scriptsize{O}$S$\scriptsize{AM}$-2.0: a tool for automated one-loop calculations within the Standard Model and beyond,''
  Eur.\ Phys.\ J.\ C {\bf 74}, no. 8, 3001 (2014)
  [arXiv:1404.7096 [hep-ph]].
  
  \bibitem{Frixione:2008ym} 
  S.~Frixione and B.~R.~Webber,
  ``The MC and NLO 3.4 Event Generator,''
  arXiv:0812.0770 [hep-ph].

  
\bibitem{Gleisberg:2008ta} 
  T.~Gleisberg, S.~Hoeche, F.~Krauss, M.~Schonherr, S.~Schumann, F.~Siegert and J.~Winter,
  ``Event generation with SHERPA 1.1,''
  JHEP {\bf 0902}, 007 (2009)
  [arXiv:0811.4622 [hep-ph]].

  
\bibitem{Alwall:2014hca} 
  J.~Alwall {\it et al.},
  ``The automated computation of tree-level and next-to-leading order differential cross sections, and their matching to parton shower simulations,''
  JHEP {\bf 1407}, 079 (2014)
  [arXiv:1405.0301 [hep-ph]].  

 
\bibitem{Passarino:2001wv}
  G.~Passarino,
  ``An Approach toward the numerical evaluation of multiloop Feynman diagrams,''
  Nucl.\ Phys.\ B {\bf 619} (2001) 257
  [hep-ph/0108252].

\bibitem{Anastasiou:2007qb}
  C.~Anastasiou, S.~Beerli and A.~Daleo,
  ``Evaluating multi-loop Feynman diagrams with infrared and threshold singularities numerically,''
  JHEP {\bf 0705} (2007) 071
  [hep-ph/0703282].

\bibitem{Becker:2012bi}
  S.~Becker and S.~Weinzierl,
  ``Direct numerical integration for multi-loop integrals,''
  Eur.\ Phys.\ J.\ C {\bf 73} (2013) 2321
  [arXiv:1211.0509 [hep-ph]].
  
\bibitem{Mahlon:1993fe} 
  G.~Mahlon,
  Phys.\ Rev.\ D {\bf 49}, 2197 (1994)
  doi:10.1103/PhysRevD.49.2197
  [hep-ph/9311213].
  
  
\bibitem{Ossola:2007bb} 
  G.~Ossola, C.~G.~Papadopoulos and R.~Pittau,
  JHEP {\bf 0707}, 085 (2007)
  doi:10.1088/1126-6708/2007/07/085
  [arXiv:0704.1271 [hep-ph]].
  
  
\bibitem{Bernicot:2007hs} 
  C.~Bernicot and J.-P.~Guillet,
  JHEP {\bf 0801}, 059 (2008)
  doi:10.1088/1126-6708/2008/01/059
  [arXiv:0711.4713 [hep-ph]].
  
\bibitem{Binoth:2007ca} 
  T.~Binoth, G.~Heinrich, T.~Gehrmann and P.~Mastrolia,
  Phys.\ Lett.\ B {\bf 649}, 422 (2007)
  doi:10.1016/j.physletb.2007.04.032
  [hep-ph/0703311].
  
\bibitem{Hahn:2004fe}
  T.~Hahn,
  ``CUBA: A Library for multidimensional numerical integration,''
  Comput.\ Phys.\ Commun.\  {\bf 168} (2005) 78
  [hep-ph/0404043].
  
\bibitem{Cuhre1}
  J.~Berntsen, T.~O.~Espelid, A.~Genz,
  ''An Adaptive Algorithm for the Approximate Calculation of Multiple Integrals''
  ACM Trans. Math. Softw. {\bf 17} (1991) 437-451.
  
\bibitem{Cuhre2}
  J.~Berntsen, T.~O.~Espelid, A.~Genz,
  ''An Adaptive Multidimensional Integration Routine for a Vector of Integrals''
  ACM Trans. Math. Softw. {\bf 17} (1991) 452-456.  

\bibitem{Lepage:1980dq}
  G.~P.~Lepage,
  ``Vegas: An Adaptive Multidimensional Integration Program,''
  Report No CLNS-80/447.
    
\bibitem{Hahn:1998yk} 
  T.~Hahn and M.~Perez-Victoria,
  ``Automatized one loop calculations in four-dimensions and D-dimensions,''
  Comput.\ Phys.\ Commun.\  {\bf 118}, 153 (1999)
  [hep-ph/9807565].

  
\bibitem{Borowka:2015mxa}
  S.~Borowka, G.~Heinrich, S.~P.~Jones, M.~Kerner, J.~Schlenk and T.~Zirke,
  ``SecDec-3.0: numerical evaluation of multi-scale integrals beyond one loop,''
  arXiv:1502.06595 [hep-ph].






\end{thebibliography}
\end{document}